\documentclass[amsmath,epsf,eclepsf,epsfig,subfigure,preprint,aps,graphicx,amssymb]{revtex4}
\usepackage{epsfig}
\begin{document}
\title{ Mass eigenstates and mass eigenvalues of seesaw$^*$}
\author{
Takuya Morozumi
\footnote{morozumi@hiroshima-u.ac.jp.  
The paper is prepared as a contribution to
the Symposium in Honour of Professor Gustavo C. Branco, CP
Violation and Flavour Puzzle.}
\bigskip\bigskip}
\address{
Graduate School of Science, Hiroshima University,
 Higashi-Hiroshima, Japan, 739-8526}
\begin{abstract}
The light neutrino mass spectrum and mixing matrix
of seesaw model including three 
right-handed neutrinos are studied for the most general case.
An approximate formulae for mass eigenvalues, mixing matrix, and CP
violation of neutrino oscillations are given.
\end{abstract}
\maketitle
\def\be{\begin{equation}}
\def\ee{\end{equation}}
\def\bea{\begin{eqnarray}}
\def\eea{\end{eqnarray}}
\def\ben{\begin{enumerate}}
\def\een{\end{enumerate}}
\def\nn{\nonumber}
\def\lsl{ l \hspace{-0.45 em}/}
\def\ksl{ k \hspace{-0.45 em}/}
\def\qsl{ q \hspace{-0.45 em}/}
\def\psl{ p \hspace{-0.45 em}/}
\def\ppsl{ p' \hspace{-0.70 em}/}
\def\dsl{ \partial \hspace{-0.45 em}/}
\def\Dsl{ D \hspace{-0.55 em}/}
\def\matrix{ \left(\begin{array} \end{array} \right) }
\def\ma{m_A}
\def\mf{m_f}
\def\mz{m_Z}
\def\mw{m_W}
\def\ml{m_l}
\def\ms{m_S}
\bibliographystyle{plain}
\thispagestyle{empty}
\section{Introduction}
It is my pleasure to write a small contribution for Professor Gustavo
fest. I visited him at Lisbon when I was a postdoctoral fellow of
Rockefeller University. Since then, I learned lots from Gustavo and his
colleagues about 
SU(2) singlet quark model \cite{branco}, and the diagonalization of
mass matrix etc. So it is appropriate 
for me to write on the neutrino masses for seesaw model
in which SU(2) singlet heavy neutral leptons are introduced.
\section{Mass eigenvalue equation for light neutrinos}
In seesaw \cite{Minkowski,Yana,Gellmann,MohaSen}, 
the effective mass term for light neutrinos is given by
the famous formulae,
$m_{eff}=-m_D \frac{1}{M} m_D^T$.
What I want to talk is about mass eigenvalues of $m_{eff}$.
It is more convenient to work for Hermite matrix, $H=m_{eff}
m_{eff}^{\dagger}$.
By denoting $\lambda$ as mass eigenvalues squared, the eigenvalue
equation becomes,
\bea
{\rm det} (H- \lambda)=0.
\eea
The equation is a qubic equation which is given by,
\bea
\lambda^3- \lambda^2 a  + \lambda b -c =0,
\eea
where,
\bea
a&=&{\rm Tr} H=H_{11}+H_{22}+ H_{33}, \nn \\
b&=& H_{11} (H_{22} H_{33} -|H_{23}|^2)
 +
H_{22} (H_{33} H_{11} -|H_{13}|^2) 
+H_{33} (H_{11} H_{22} -|H_{12}|^2),
\nn \\
c&=& {\rm det} H.
\eea
The coefficients are related to mass squared eigenvalues $n_1^2, n_2^2, n_3^2$
as,
\bea
a&=&n_3^2+n_2^2+n_1^2 \equiv 3 {\bar n}^2\nn \\
b&=& n_3^2 n_2^2 +n_1^2 n_2^2+ n_2^2 n_3^2 \nn \\
 &=&  3 {\bar n}^4 + (n_1^2-\bar{n}^2)(n_2^2-\bar{n}^2)
 +(n_2^2-\bar{n}^2)(n_3^2-\bar{n}^2)
 (n_3^2-\bar{n}^2)(n_1^2-\bar{n}^2), \nn \\
c&=& n_3^2 n_2^2 n_1^2 = \bar{n}^6+(n_1^2-\bar{n}^2)
(n_2^2-\bar{n}^2)(n_3^2-\bar{n}^2) +\bar{n}^2 [
(n_2^2-\bar{n}^2) (n_3^2-\bar{n}^2)\nn \\
&+&(n_3^2-\bar{n}^2)(n_1^2-\bar{n}^2)
+(n_1^2-\bar{n}^2)(n_2^2-\bar{n}^2)].
\eea
It is interesting to see except $\bar{n}^2$, 
$n_i^2-\bar{n}^2$ $(i=1,2,3)$ can be written in terms of 
the mass squared differences which are measured in oscillation
experiments.
Now let us write each coefficient $a,b$ and $c$ in terms
of the element of $m_D$ and $M$ explicitly.
Without loss of generality, we can take $m_D$ is a general 
$3 \times 3$ complex matrix and $M$ is a $3 \times 3$ real diagonal
matrix.
We introduce a decomposition. \cite{Fujihara}
\bea
&&m_D=({\bf m_{D1}}, {\bf m_{D2}}, {\bf m_{D3}})
=({u_1},{u_2},{u_3}) \left(
\begin{array}{ccc} m_{D1} & 0 & 0 \\
                    0 & m_{D2} & 0 \\
                    0 & 0 & m_{D3} \end{array} \right) \nn \\
&&=U \times {\rm diagonal}(m_{D1}, m_{D2}, m_{D3}).
\eea
where $\frac{\bf m_{Di}}{m_{Di}}={u_i}$ and $|{u_i}|=1$ with $m_{Di}=
|{\bf m_{Di}}|$. $u_i (i=1 \sim 3) $ is a complex vector in $C^3$.
$U=({u_1},{u_2},{u_3}) $ is not unitary in general.
\bea
A \equiv U^{\dagger} U= \left(\begin{array}{ccc}
1 & { u_1^{\dagger} \cdot u_2} & {u_1^{\dagger} \cdot u_3} \\
\ast & 1  & { u_2^{\dagger} \cdot u_3} \\
\ast & \ast & 1 \end{array} \right),
\eea
where $A$ is an Hermite matrix.
One can write,
\bea
m_{eff}&=& -U X U^T, \nn \\
H&=&m_{eff} m_{eff}^{\dagger}=U X A^{\ast} X U^{\dagger},
\eea
where $X$ is a real diagonal matrix with mass dimension,
\bea
X={\rm Diagonal} \left(X_1,X_2,X_3 \right) \
{\rm with} \ X_i \equiv \frac{m_{Di}^2}{M_i}.
\eea
Using the definitions, one may write the coefficients of cubic equation
as,
\bea
a&=&{\rm Tr} (A X A^{\ast} X)=X_1^2+X_2^2+X_3^2\nn \\
&+&2 {\rm Re}(A_{12}^2) X_1 X_2 +
2 {\rm Re}(A_{23}^2) X_2 X_3 +
2 {\rm Re}(A_{31}^2) X_3 X_1,
\nn \\
b&=& X_1^2 X_2^2 \left(1-|A_{12}|^2\right)^2+
X_2^2 X_3^2\left(1-|A_{23}|^2\right)^2+X_3^2 X_1^2
\left(1-|A_{31}|^2\right)^2 \nn \\
&+&2 X_1 X_2 X_3^2 
{\rm Re}\left((A_{12}-A_{13} A_{32})^2 \right)
+ 2 X_2 X_3 X_1^2 
{\rm Re}\left((A_{23}-A_{21} A_{13})^2 \right) \nn \\
&+& 2 X_3 X_1 X_2^2 
{\rm Re}\left((A_{31}-A_{32} A_{21})^2 \right), \nn \\
c&=& ({\rm det} A)^2 X_1^2 X_2^2 X_3^2 \nn \\
 &=& \left(1+2{\rm Re} (A_{12} A_{23} A_{31}) -|A_{12}|^2-|A_{23}|^2-|A_{31}|^2
 \right)^2
     X_1^2 X_2^2 X_3^2.
\eea
In this paper, we discuss the case with $X_1 \ll X_2 \ll X_3$ and
$|A_{23}| < 1$.
If we neglect $X_1$, the eigenvalue equation becomes quadratic as,
\bea
\lambda^2-(X_2^2+X_3^2 + 2 {\rm Re}(A_{23}^2) X_2 X_3) \lambda + X_2^2 X_3^2
(1-|A_{23}|^2)^2=0.
\eea
Therefore, the heaviest mass squared $n_3^2$ 
at leading order is given as,
\bea
n_3^2=X_3^2.
\eea
Then, the smaller eigenvalue is given as,
\bea
n_2^2=X_2^2(1-|A_{23}|^2)^2.
\eea
Finally the mass squared of the lightest state is,
\bea
n_1^2=X_1^2 \frac{({\rm det A})^2}{(1-|A_{23}|^2)^2}.
\eea
\section{mixing matrix}
Now let us turn to the mixing matrix which reroduces the
approximate eigenvalues given in previous sections.
One may first write the $m_{eff}$ as
\bea
m_{eff}=-X_3 ({u_1, u_2, u_3}) {\rm Diag}.(X_1/X_3,X_2/X_3,1) (
{ u_1,u_2,u_3})^T.
\eea
By taking the limit  $X_1/X_3 \rightarrow 0$, 
$m_{eff}$ 
becomes, 
\bea
m_{eff}=-X_3 (u_2 \frac{X_2}{X_3} u_2^T + u_3 u_3^T).
\eea
$m_{eff}$ has a zero eigenvalue and the corresponding state 
can be isolated by using a unitary rotation given as,
\bea
V_0&=&(v_1, v_2, v_3), \nn \\
v_1^{\dagger}\cdot u_2&=&v_1^{\dagger} \cdot u_3=0, \nn \\
v_2^{\dagger} \cdot u_3&=&0, \nn \\
v_3 &\equiv& u_3.
\eea
One may write,
\bea
v_2&=&\frac{u_2-u_3 A_{32}}{\sqrt{1-|A_{32}|^2}}, \nn \\
v_1&=& \frac{u_2^{\ast} \times u_3^{\ast}}{\sqrt{1-|A_{32}|^2}}.
\eea
Using the definition, one may show,
\bea
\lim_{X_1 \rightarrow 0} V_0^{\dagger} m_{eff} V_0^{\ast}
=-X_3 \left(\begin{array}{ccc}
             0 & 0 & 0 \\
             0 & \frac{X_2}{X_3}(1-|A_{23}|^2) & \frac{X_2}{X_3} A_{32} \sqrt{
             1-|A_{23}|^2} \\
             0 & \frac{X_2}{X_3} A_{32} \sqrt{1-|A_{23}|^2} & A_{32}^2 
             \frac{X_2}{X_3} +1 \end{array} \right).
\eea
The final step of the diagonalization
can be achieved by diagonalizaing two by two sector.
We write MNS (Maki, Nakagawa and Sakata) \cite{MNS} matrix V as,
$V=V_0 K$ and $K$ is given as,
\bea
K=\left( \begin{array}{ccc} 
        1 & 0 & 0 \\
        0 & \cos \theta_N & \sin \theta_N \exp(-i \phi_N) \\
        0 & -\sin \theta_N \exp(i \phi_N)
        & \cos \theta_N \end{array}  \right) P,
\eea
where $P$ is a diagonal phase matrix which explicit form is given in
Eq.(38) of \cite{Fujihara}. $\tan 2 \theta_N $ 
is also obtained by simply replacing
$X_1 \rightarrow X_3$ and $u_1 \rightarrow u_3$ in 
Eq.(38) of \cite{Fujihara}.
Explicitly, it is given as,
\bea
\tan 2 \theta_N&=& \frac{2 X_2 \sqrt{1-|A_{23}|^2} |X_3 A_{23} +X_2 A_{32}|}
{X_3^2 + X_2^2(2 |A_{23}|^2-1)+ 2 X_2 X_3 Re.(A_{23})^2}. 
\eea
When $X_2 < X_3$,
\bea
\theta_N \simeq \frac{X_2}{X_3} \sqrt{1-|A_{23}|^2}|A_{23}| \ll 1.
\eea
Therefore in the leading order of $X_2/X_3$ expansion,
$V=V_0$.
\section{CP violation of neutrino oscillation}
It would be interesting to see how CP violation of neutrino 
oscillation looks like. One may compute CP violation of neutrino
oscillation \cite{branco2} at the leading power
of the term of $X_1^n X_2^m X_3^{6-m-n}$.\cite{Fujihara}
 One find the term
which is proportional to $X_2^2 X_3^4$
is the leading and obtain,
\bea
\Delta=&& {\rm Im} \left(
(m_{eff} m_{eff}^{\dagger})_{e \mu} (m_{eff} m_{eff}^{\dagger})
_{\mu \tau} (m_{eff} m_{eff}^{\dagger})_{\tau e} \right) \nn \\
&& =(1-|A_{23}|^2)
(X_2^2 X_3^4)
({\rm Im}[(u_{e3}^{\ast} u_{e2} u_{\mu3} u_{\mu2}^{\ast})
|u_{\tau3}|^2+ \nn \\
&& {\rm Im}[(u_{\mu3}^{\ast} u_{\mu 2} u_{\tau3} u_{\tau2}^{\ast})
|u_{e 3}|^2+ {\rm Im}[(u_{\tau 3}^{\ast} u_{\tau 2} u_{e 3} u_{e 2}^{\ast})
|u_{\mu 3}|^2).
\eea
As a check of the result, we may compute the Jarlskog invariant 
and mass squared difference separately and obtain $\Delta$.
\bea
&& \Delta=J (n_1^2-n_2^2) (n_2^2-n_3^2) (n_3^2-n_1^2),\nn \\
&&(n_1^2-n_2^2)(n_2^2-n_3^2)(n_3^2-n_1^2)\simeq n_2^2 n_3^4
\simeq X_3^4 X_2^2 (1-|A_{23}|^2)^2,\nn \\
&&J=-{\rm Im}{V_{e3} V_{\mu 3}^{\ast} V_{e2}^{\ast} V_{\mu2}} \simeq
 -{\rm Im} \left(
 u_{e3} u_{\mu 3}^{\ast} \frac{(u_{e2}^{\ast}-u_{e3}^{\ast}A_{23})
      (u_{\mu2}-u_{\mu3} A_{32})}{1-|A_{23}|^2} \right),\nn \\
\eea
which agrees with $\Delta$ computed previously.
\section{Summary}
We have studied eigenvalues and mixing matrix of seesaw model
with three right-handed neutrinos. The most general eigenvalue
equation which determines the light neutrino mass spectrum is given.
The equation is solved by assuming the hierarchy of the three parameters
$X_i (i=1 \sim 3)$ as $X_1 < X_2 < X_3$. Under the assumption,
the analysis of the model with two right-handed neutrino
\cite{Fujihara} is useful.
 The leading order expression of mass spectrum
and MNS matrix is obtained in terms of $X_i$ and normalized
yukawa vectors $u_i$. We compute the leading term of CP violation
of neutrino oscillations which is a product of 
the Jarlskog invariant and  mass differences.
All the leading contribution can be extracted in analytical 
form which may be useful for further study on CP violation 
at low energy and leptogenesis. \cite{FY} \cite{Fujihara,
branco2,FGY,EKKMT,EMOP}
\def\apj#1#2#3{Astrophys.\ J.\ {\bf #1}, #2 (#3)}
\def\ijmp#1#2#3{Int.\ J.\ Mod.\ Phys.\ {\bf #1}, #2 (#3)}
\def\mpl#1#2#3{Mod.\ Phys.\ Lett.\ {\bf A#1}, #2 (#3)}
\def\nat#1#2#3{Nature\ {\bf #1}, #2 (#3)}
\def\npb#1#2#3{Nucl.\ Phys.\ {\bf B#1}, #2 (#3)}
\def\plb#1#2#3{Phys.\ Lett.\ {\bf B#1}, #2 (#3)}
\def\prd#1#2#3{Phys.\ Rev.\ {\bf D#1}, #2 (#3)}
\def\ptp#1#2#3{Prog.\ Theor.\ Phys.\ {\bf #1}, #2 (#3)}
\def\pr#1#2#3{Phys.\ Rev.\ {\bf #1}, #2 (#3)}
\def\prl#1#2#3{Phys.\ Rev.\ Lett.\ {\bf #1}, #2 (#3)}
\def\prp#1#2#3{Phys.\ Rep.\ {\bf #1}, #2 (#3)}
\def\sjnp#1#2#3{Sov.\ J.\ Nucl.\ Phys.\ {\bf #1}, #2 (#3)}
\def\zp#1#2#3{Z.\ Phys.\ {\bf #1}, #2 (#3)}
\def\jhep#1#2#3{JHEP\ {\bf #1}, #2 (#3)}
\def\epjc#1#2#3{Euro. Phys. J.\ {\bf C#1}, #2 (#3)}
\def\rmp#1#2#3{Rev. Mod. Phys.\ {\bf #1}, #2 (#3)}
\def\prgth#1#2#3{Prog. Theor. Phys.\ {\bf #1}, #2 (#3)}
\subsection{Acknowledgement}
The author thanks Prof. M. N. Rebelo and organizers of Gustavofest
for giving me a chance for writing contribution. He thanks T. Fujihara,
S. K. Kang, C.S. Kim and D. Kimura for discussion. The work is
supported by
the kakenhi, Japan, No.16028213.


\begin{thebibliography}{99}
\bibitem{branco}G.C. Branco, T. Morozumi, P. A. Parada, and M. N. Rebelo,
\prd{48}{1167}{1993}.
\bibitem{Minkowski}
P. Minkowski,
\plb{67}{421}{1977}.
\bibitem{Yana}
T. Yanagida, in the proceedings of the Workshop
on Unified Theories and Baryon Number in the Universe,
edited by O. Sawada and A. Sugamoto, 95 (1979).
\bibitem{Gellmann}
M. Gell-Mann, P. Ramond and R. Slansky,
in Supergravity, P. van Nieuwenhuizen and D.Z. Freedman
(eds.), North Holland Publ. Co.,(1979).
\bibitem{MohaSen}
R. N. Mohapatra and G. Senjanovich,
\prl{44}{912}{1980}.
\bibitem{Fujihara} T. Fujihara, S. Kaneko, S. Kang, D. Kimura,
T. Morozumi and
M. Tanimoto
\prd{72}{016006}{2005}.
\bibitem{MNS} Z. Maki, M. Nakagawa, and S. Sakata,
\ptp{28}{1174}{1962}.
\bibitem{branco2} G. C. Branco, T. Morozumi, B. Nobre, and M. N. Rebelo
\npb{617}{475}{2001}.
\bibitem{FY} M. Fukugita and T. Yanagida, \plb{174}{45}{1986}.
\bibitem{FGY} P.H. Frampton, S.L. Glashow, and T. Yanagida,
\plb{548}{119}{2002}.
\bibitem{EKKMT} T. Endoh, S. Kaneko, S. K. Kang, T. Morozumi and M. Tanimoto,
\prl{89}{231601}{2002}.
\bibitem{EMOP} T. Endoh, T. Morozumi, A. Purwanto and T. Onogi,
\prd{64}{013006}{2001}, Erratum-ibid.D64:059904,2001.
\end{thebibliography}
\end{document}